%% file: 49P.tex
\documentclass[twocolumn]{aastex62}
\title{49P Paper}

\newcommand\aastex{AAS\TeX}

\usepackage{xcolor}
\usepackage{url}
\usepackage{longtable} 
\usepackage{times}
\usepackage{float}
\usepackage{wrapfig}
\usepackage{rotating}
\usepackage{amssymb}
\usepackage{xcolor}

\def\hawaii{Hawai$\!$`i}

\received{March 20, 2019}
\accepted{November 1, 2019}

\submitjournal{Icarus}

\shorttitle{\aastex\ 49P/Arend-Rigaux}
\shortauthors{Chu et al.}

\begin{document}

\title{Detailed Characterization of Low Activity Comet 49P/Arend-Rigaux}

\correspondingauthor{Laurie E. U. Chu}
\email{lurban@ifa.hawaii.edu}

\author[0000-0002-1437-4463]{Laurie E. U. Chu}
\affiliation{Institute for Astronomy, 640 N. Auhoku Pl. \#209, Hilo, HI 96720 USA}

\author[0000-0002-2058-5670]{Karen J. Meech}
\affiliation{Institute for Astronomy, 2680 Woodlawn Drive, Honolulu, HI 96822 USA}

\author[0000-0002-4767-9861]{Tony L. Farnham}
\affiliation{Department of Astronomy, University of Maryland, College Park, MD 20742-2421}

\author{Ekkehard K{\"u}hrt}
\affiliation{DLR German Aerospace Center, Institute of Planetary Research, Rutherfordstr. 2, D-12489 Berlin., Germany}

\author[0000-0002-0457-3872]{Stefano Mottola}
\affiliation{DLR German Aerospace Center, Institute of Planetary Research, Rutherfordstr. 2, D-12489 Berlin., Germany}

\author[0000-0002-2021-1863]{Jacqueline V. Keane}
\affiliation{Institute for Astronomy, 2680 Woodlawn Drive, Honolulu, HI 96822 USA}

\author{Stephan Hellmich}
\affiliation{DLR German Aerospace Center, Institute of Planetary Research, Rutherfordstr. 2, D-12489 Berlin., Germany}

\author[0000-0001-6952-9349]{Olivier R. Hainaut}
\affiliation{European Southern Observatory, Karl-Schwarzschild-Strasse 2, D-85748 Garching bei M\"unchen, Germany}

\author[0000-0002-4734-8878]{Jan T. Kleyna}
\affiliation{Institute for Astronomy, 2680 Woodlawn Drive, Honolulu, HI 96822 USA}

\begin{abstract}
Comet 49P/Arend-Rigaux is a well known low-activity Jupiter Family comet.  Despite the low activity, we have witnessed outgassing activity in 1992, 2004, and 2012.  In 2012 a broad tail-like feature (PA$\sim$270$^\circ$, $\sim$2.3$\times$10$^5$ km) and a narrow jet-like feature (PA$\sim$180$^\circ$, $\sim$9.3$\times$10$^4$ km) were seen simultaneously.  Using Finson-Probstein (FP) dust dynamical models we determine: the grain sizes released in each event; the duration of activity; when the activity peaked; and the velocity of the dust particles, allowing us to make comparisons between the events.  We find that the tail feature in 2012 is similar to the tail in 1992 with large grains (40-4000 $\micron$) peaking in activity around perihelion with a long duration of outgassing greater than 150 days.  The jet feature from 2012, however, is more similar to the 2004 event which we model with small grains (1-8 $\micron$) with a short duration of activity on the order of one month.  The main difference between these two features is that the 2004 event occurs prior to perihelion, while the 2012 event is post-perihelion.  We use the grain sizes from the FP models to constrain ice sublimation models.  Between 1985 and 2018 we cover six apparitions with 26 nights of our own observations, data from the literature, and data from the Minor Planet Center, which together, allow us to model the heliocentric light curve.  We find that the models are consistent with H$_2$O ice sublimation as the volatile responsible for driving activity over most of the active phases and a combination of H$_2$O and CO$_2$ ices are responsible for driving activity near perihelion.  We measure the fractional active area over time for H$_2$O and discover that the activity decreases from an average active area of $\sim3\%$ to $\sim0.2\%$.  This clear secular decrease in activity implies that the comet is becoming depleted of volatiles and is in the process of transitioning to either a dormant or dead state. 

\end{abstract}



\section{Introduction}\label{sec:intro}

The presence of volatiles in comets distinguishes them from asteroids. It may be that some asteroids were once comets that ran out of volatiles, became dormant, or have such a low activity that they are classified as asteroids.  The transition to an asteroidal state may come from the cracking and resealing of the mantle causing intermittent cometary activity \citep{jewitt2004}. From the Rosetta Mission's observations of 67P/Churyumov-Gerasimenko, dust emitted during an active stage may fall back to the comet and trap some of the ices beneath the surface, which also could contribute to a change in the activity state \citep{Hu2017}.  Comet 49P/Arend-Rigaux (hereafter, 49P) is suspected to be at this stage.

Comet 49P/Arend-Rigaux was discovered in 1951 and is a member of the Jupiter Family comets (JFC, orbital period less than 20 years and an inclination less than $\sim$30$^{\circ}$; dynamically controlled by Jupiter). It has an orbital period of 6.72 years, with perihelion at $q$=1.42 au and aphelion at $Q$=5.70 au  (2011 Epoch).   The orbit has been very stable; it has not passed within 0.9 au of Jupiter for 900 years or more \citep{Marsden1970}. The comet has a relatively large nucleus of $R_N$=4.60$\pm$0.11 km and an albedo p$_V$=0.03, typical of most comets \citep{lowry2003, millis1988}.  Its small perihelion distance and large size makes 49P/Arend-Rigaux relatively bright and easy to observe. Because of the low activity, this was one of  the first comets for which a rotation light curve was measured \citep{jewitt1985, wisniewski1986, millis1988}. The rotation period is reported to be 13.47$\pm$0.017 \citep{millis1988})  and 13.45$\pm$0.005 hr \citep{eisner2017} with a brightness that varies up to 0.4 magnitudes as the comet rotates \citep{jewitt1985}. 

Since its discovery, 49P has characteristically displayed low activity even when near perihelion. In December 1998 when it was at $r$=2.11 au, \citet{lowry2001} observed the comet and noted an almost stellar appearance except for a small dust-jet emanating in the anti-solar direction. They suggested this could be due to pressure buildup near areas of thin mantle resulting in temporary outbursts, and that the comet should be monitored for any possible transition to an asteroid-like state. Observations with the {\it Spitzer Space Telescope} did not show a debris trail, consistent with it being a minimally active comet without large dust particles \citep{reach2007}.

We observed 49P/Arend-Rigaux near perihelion in 2012 as it was outward bound. Despite its low activity, we witnessed both a tail feature pointing in the anti-solar direction at a position angle (PA) of $\sim$270$^\circ$ and a narrow dust-jet pointing at PA$\sim$180$^\circ$.  Because of its known low activity these features were unexpected, which leads us to consider if this comet is becoming more asteroidal or if a volatile-rich reservoir remains.  We have a long baseline of 49P/Arend-Rigaux observations covering several apparitions which allows us to explore the activity and any secular changes, and probe what drives the activity in this comet.  

\input{tab-photom_new.tex}

\section{Observations and Data Collection\label{sect:observations}}

\begin{figure}[h!t]
\centering
\includegraphics[width=0.46\textwidth]{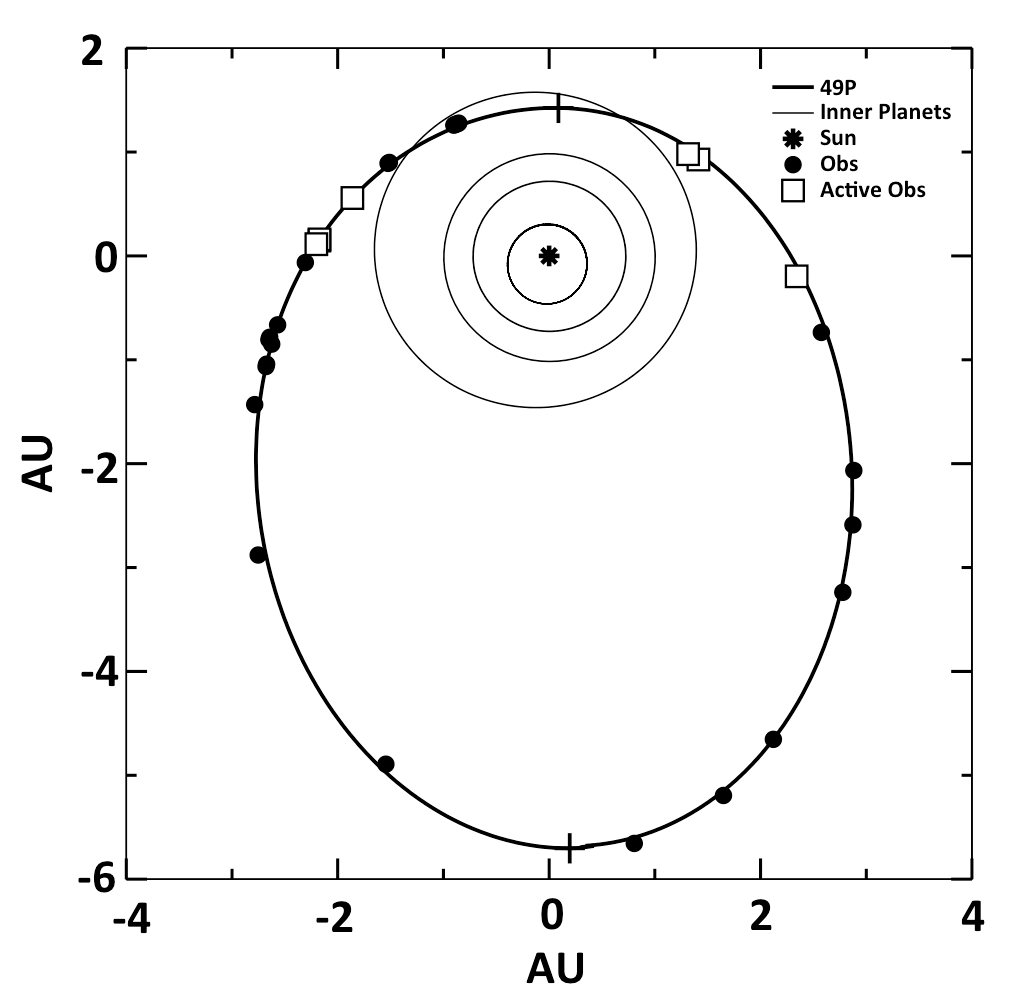}
\caption{Orbit plane diagram of 49P/Arend-Rigaux showing the positions along the orbit at which we have data from Table~\ref{tab:photom}  (including supplemental literature observations). The orbits of the inner planets are drawn for reference. Perihelion and aphelion are marked with a dash on the 49P orbit. The observations marked active showed either a jet or a tail in the images, or gas was detected spectroscopically.}
\label{fig:orbit}
\end{figure}

We have 26 nights of observations from 1985 to 2012 covering five apparitions (Figure~\ref{fig:orbit}). The telescope, instruments, and the observing geometry for each date of observation are provided in Table~\ref{tab:photom}.  We make use of additional data taken from the literature from five nights (Figure~\ref{fig:orbit}), and the same information for these observations are reported in Table~\ref{tab:photom}.

\subsection{Observation of Activity}

\begin{figure}[h!tb]
\centering
\includegraphics[scale=0.16]{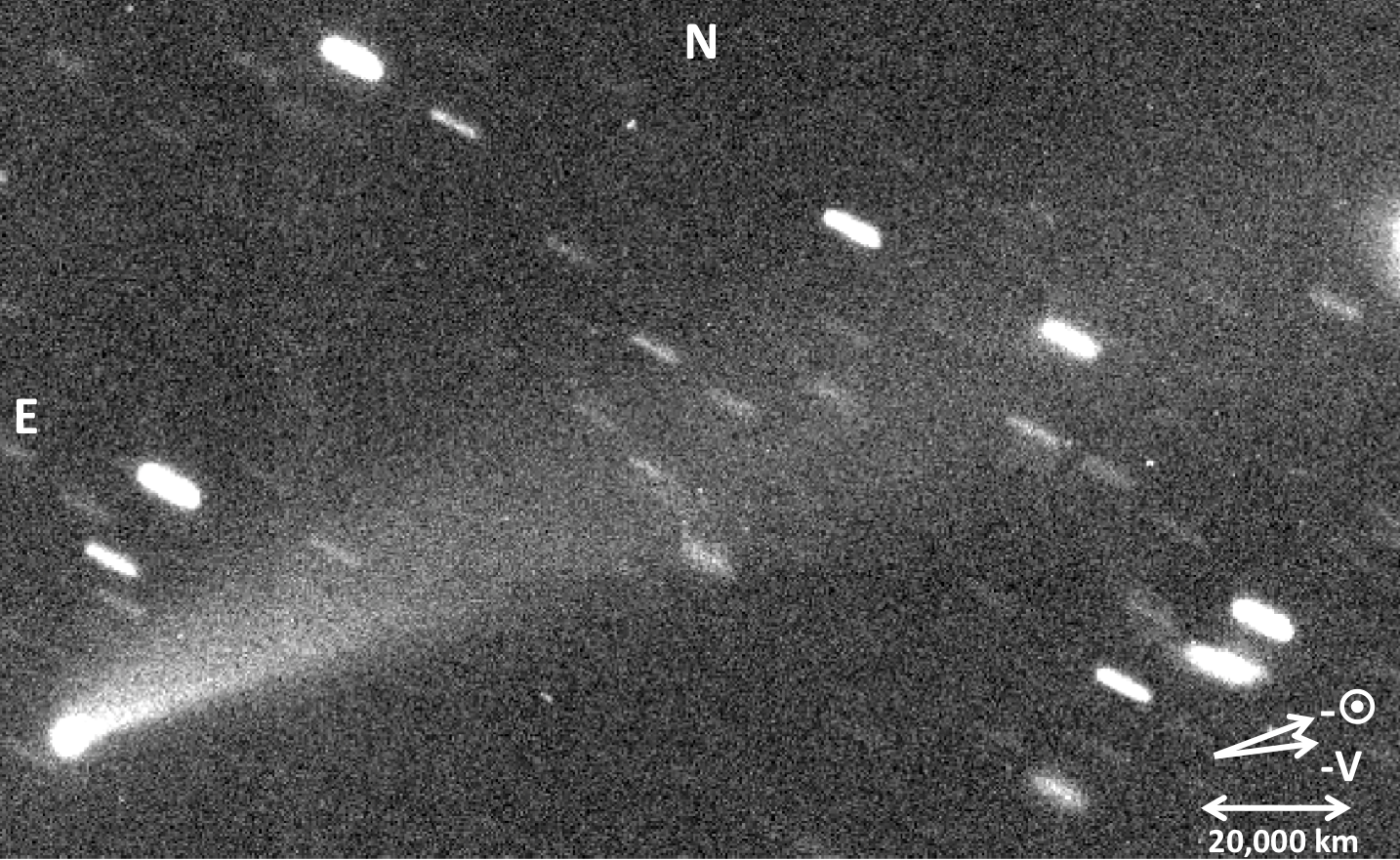}
\caption{An individual image taken of 49P/Arend-Rigaux on Jan 6, 1992 taken using the UH 2.2m telescope on Maunakea.  The image is 3.36\arcmin $\times$ 2.05\arcmin.  This image was used for modeling with the Finson-Probstein models (see discussion in section 6.1.1).}
\label{fig:jan1992}
\end{figure}

\begin{figure}[h!tb]
\centering
\includegraphics[scale=0.16]{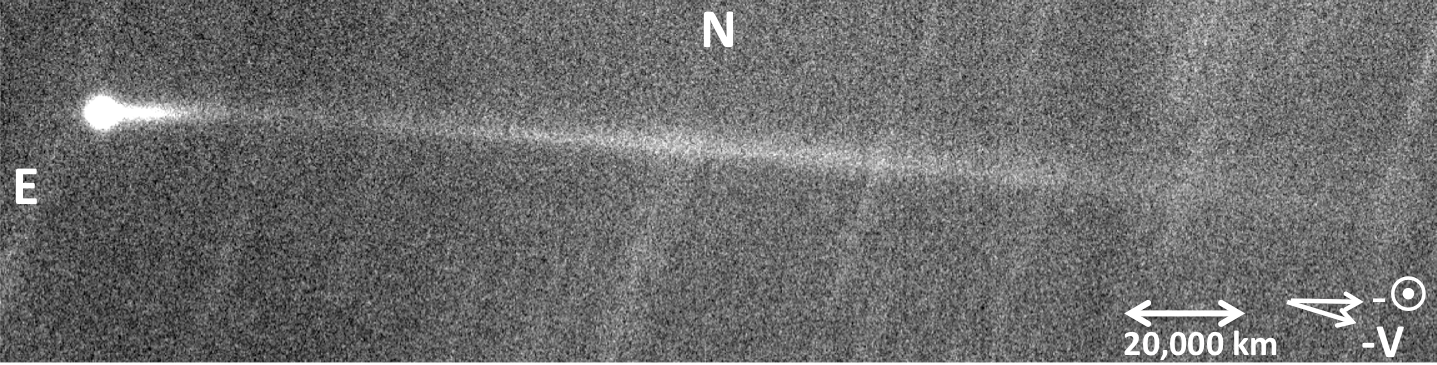}
\caption{Stacked image of 49P/Arend-Rigaux on August 23, 2004 from \citet{stevenson2007}. The image is 3.6\arcmin $\times$ 0.9\arcmin. The jet-like feature at PA$\sim$270$^\circ$is similar to the feature seen in the 2012 images and modeled using Finson-Probstein models shown in Figure \ref{fig:FPmodels}. }
\label{fig:aug2004}
\end{figure}

For the majority of the observations, comet 49P appeared stellar and did not show any apparent activity.  The first noticeable dust emission in our data set appeared in 1992 (after perihelion) as a dust tail with two features. The projected length of the main tail was $\sim$1.7$\times$10$^5$ km and was directed at a PA$\sim$295$^\circ$.  A small narrow tail extended with a projected length of $\sim$6$\times$10$^4$ km in the anti-solar direction with PA$\sim$292$^\circ$.  An image showing this activity is in Figure~\ref{fig:jan1992}.  During a later apparition, in 2004, \citet{stevenson2007} present data showing a very narrow jet-like feature that extended from the surface at a PA$\sim$270$^\circ$, 186 days pre-perihelion (Figure~\ref{fig:aug2004}).  \citet{reach2007, reach2013} found that 49P was actively producing CO$_2$ during the same apparition in 2004.  

In 2012 we observed a tail and jet feature similar to the one seen by \citet{stevenson2007}.  Using data from amateur archives\footnote{available at http://comet.observations.free.fr} we know that there was a tail pointing in the antisolar direction with PA$\sim$270$^\circ$ starting as early as 46 days before perihelion (which occurred on October 19, 2011).  Our first observations during this apparition of 49P were on February 22, 2012, where we still saw a tail with a projected length of $\sim$2.3$\times10^5$ km. We observed 49P later on March 28 and saw another feature at PA$\sim$180$^\circ$ that resembled a narrow jet. The tail was also still present.  Using the amateur archives we discovered that the jet appeared some time between March 16 and March 23. The jet extended $\sim$9.3$\times10^4$ km (projected length) and was faint close to the comet, then had a small bright patch, and then faded again further from the nucleus. We followed up with observations on April 1, when the jet remained visible but the bright portion had moved slightly away from the nucleus.  The tail was not noticeably different.  We observed 49P again on April 15 when only a small feature appeared close in to the comet pointing at PA$\sim$180$^\circ$ and a faint tail remained. By June 5, the nucleus was stellar in appearance and the tail was no longer seen. This sequence is shown in Figure \ref{fig:FPmodels}.  With several instances of activity we seek to determine whether this comet has depleted most of its volatile supply or if there remains a rich deep reservoir of ices that causes these outbursts.    

\section{Data Reduction and Analysis}

Out of the 26 nights of data used, four nights were already fully processed by \citet{jewitt1985}, and the remaining 22 nights required flattening and calibrations.  For all of the data observed on the UH 2.2m, KPNO 2.1m, and CTIO 1.5m and 4m telescopes we used the Image Reduction Analysis Facility (IRAF; \citet{tody1986}) to perform bias subtraction and flat-field reductions from dithered images of the twilight sky. The images from the Calar Alto 1.2m telescope were reduced using the instrument's processing pipeline.  Non-photometric nights and the nights observed at Calar Alto were calibrated using field stars from the Sloan Digital Sky Survey (SDSS, \citet{york2000}). We then transformed the magnitudes to Kron-Cousins $R$-band equivalents using conversions derived by R. Lupton\footnote{available online at http://www.sdss.org/} and calibrated the photometry using the standards from \citet{landolt1992}.  The SDSS calibrations were tested on a photometric night where we used standard stars to determine that the results were consistent to within 3-5\%. Each of the fluxes for standard stars and field stars were measured with circular apertures that sampled the background with surrounding circular annuli.

The photometric measurements of 49P were performed using circular apertures and when the comet was active, the background sky statistics were measured manually in regions of blank sky near the object.  This was done to avoid contamination from any dust in the coma.  The aperture radius for the photometry was chosen to be at 5\farcs0 to encompass $>$99\% of the total flux from the comet while minimizing interference from nearby stars or galaxies.  The aperture size was also chosen to remain consistent across all data sets to be used for ice sublimation models discussed in Section~\ref{sect:icesub}.  Field stars in the comet images were also measured to correct for extinction variations throughout the night.

\section{Finson-Probstein Analysis}

 We use the method of Finson and Probstein \citep{finson1968a,finson1968b} to model the dust tail.  The Finson-Probstein (FP) method is a dust-dynamical method which calculates the trajectories of an ensemble of dust grains ejected from the surface of the nucleus at different times and considers the perturbations from solar radiation pressure on their orbits around the sun. A tail is modeled by summing all of the scattered light contributions from the dust grains.  By matching the synthetic dust trails to observed images it is possible to constrain the dust particle size distribution, production rate, emission velocity, and the onset and cessation of the activity. 

Original FP models were performed initially on only a few comets \citep{finson1968b,sekanina1973,jambor1973} but then the method was improved by \citet{fulle1987a} when he accounted for the time-dependence of the size distribution of particles and then refined with numerical simulations \citep{Fulle1989}.  His modified version was used in several studies \citep{fulle1987b,fulle1990,fulle1992} and to estimate dust production rates during the Giotto encounter of comet 26P/Grigg-Skjellerup \citep{fulle1993}. Others have used simplified FP models \citep{sekanina1974a, sekanina1974b, gary1974, beisser1992} to determine individual parameters such as the grain sizes, activity levels, or using streamers in the dust tail to look for rotation periods. The FP method was improved by \citet{farnham1996}  who accounted for orbital mechanic effects to allow the tail to be modeled from any comet orientation, grain size, geometry, and emission time. We use this improved model for 49P/Arend-Rigaux.

\subsection{Finson Probstein Models}

\subsubsection{Dust grain size}

We assume that there is no mass loss (from either sublimation or fragmentation) from the spherical grains leaving the nucleus. Therefore, the only forces acting on the emitted grains are the gravitational attraction from the sun and solar radiation pressure.  To a first order approximation both forces act in the radial direction and vary as $r^{-2}$ where $r$ is the heliocentric distance.  To relate these two forces a common parameter is used called $\beta$, defined as
\begin{equation}
\beta \equiv \frac{F_{rad}}{F_g}.
\end{equation}

\noindent which can be written as
\begin{equation}
\beta = 5.740 \times 10^{-4}\frac{Q_{pr}}{\rho_da}
\label{equation:beta2}
\end{equation}

\noindent 
where $a$ is the grain radius in meters, $\rho_d$ is the grain density in kg m$^{-3}$ and $Q_{pr}$ is the radiation pressure efficiency.  For our models we make certain assumptions for $\rho_d$ and $Q_{pr}$ and vary $\beta$ to match our observations. This allows us to keep the grain radius as a free parameter. We assume that the grains are spherical because without additional constraints from {\it in situ} observations from space-based data we cannot model irregular grains.

From measurements of IDPs and {\it in situ} observations of comets, we know that the minerals in the refractory dust have a typical density of $\sim$3500 kg m$^{-3}$ but we do not know the porosity, therefore the bulk density could be very different. \citet{mukai1989} and \citet{grun1990} calculate the densities of different dust materials that are possible on comets. These materials have been measured {\it in situ} and by ground-based measurements of emission features. The different materials are listed in Table~\ref{tab:dust} with the corresponding densities.  From the Rosetta mission, both fluffy and compact grains were observed \citep{Fulle2015, Hilchenbach2016} so  we take the density of 2200 kg m$^{-3}$ to use in Equation~\ref{equation:beta2} as a reasonable upper limit.

\input{tab-dust}

Detailed calculations of the radiation pressure efficiency, $Q_{pr}$, of spherical grains being illuminated by sunlight show that $Q_{pr}$ $\sim$1 and for different classes of particles the value varies by less than a factor of 2 for almost all cases \citep{burns1979}. Only metallic particles have $Q_{pr} >$2, but these particles have not been observed in cometary dust.  We adopt an average efficiency value of $Q_{pr}=1.5$. 

\subsubsection{Particle Size Distribution}

The particle size distribution can take many forms but the simplest is a differential power law as given in the equation, 
\begin{equation}
\label{equation:beta_n}
f(\beta) \propto \beta^n
\end{equation}

\noindent 
where $n$ can be as small as $\sim$1 \citep{Lisse1998}. This controls how steep the distribution is between the grain sizes and the model is generally sensitive to this parameter. The slope is controlled mostly by the particle size distribution in the nucleus but may also result from factors like the dust-to-gas ratio, the porosity of the grains, and the forces on the grains which may force small particles to join or large particles to split.  A power-law size distribution for the dust was derived from the The COmetary Secondary Ion Mass Analyzer (COSIMA) during the Rosetta Mission.  They find a form where the number of particles of a specific size, $r$, is proportional to $r^{−b}$.  The power index ranges from -1.9 for grains between 30 $\mu$m and 150 $\mu$m and -0.8 for for grains larger than 150 $\mu$m \citep{Merouane2016}.

\subsubsection{Velocity Distribution}

The initial velocity of the grains is controlled by the terminal velocity of dust in a gas flow (which is reached within a few nucleus radii), after that the grain motion is controlled by solar radiation pressure and solar gravity.  Basic orbital mechanics can be used to calculate the positions of grains at any given time.   The time at which a particle is emitted is defined as the moment the gas sublimation imparts enough energy to a dust grain that it is able to overcome the escape speed of the nucleus.  It reaches its terminal velocity and when the dust is far enough away from the nucleus it will no longer be influenced by the gravity of the nucleus or the gas flow and it begins its own orbit around the sun.  The dust grain's emission velocity is defined as the residual velocity after the gas drag (the force that pushes the grain off the nucleus) becomes negligible. The equation for the motion of a particle relates the upward force due to outflowing gas and the gravitational force of the nucleus to the motion of a dust particle.  The relation shows that if gas is emitted uniformly over the nucleus then the gas density drops off as $\rho_g \propto R^{-2}$ where $R$ is the distance from the nucleus.  Also the gravitational force will drop off as $R^{-2}$ meaning any particle that is lifted off of the surface will ultimately escape the comet.  A maximum grain size that can be lifted off of the comet can be calculated.  On the other end, very small grains would be strongly coupled to the gas flow and will approach the gas outflow velocity.  The size of grains between those that couple to the gas flow and those that cannot be lifted from the surface follow a velocity distribution using the equation:

\begin{equation}
v_i=v_{\rm max}\beta^m.
\end{equation}

\noindent  where $v_{\rm max}$ is the velocity of tiny particles ($\beta=1$), and $v_i$ is the initial emission velocity, which cannot be larger than $v_{\rm max}$ because we only allow for $\beta$ values from 0 to 1.  We use a starting value for $m$ around 0.35, but the model is relatively insensitive to the velocities so that $v_{\rm max}$ and $m$ do not need to be specified precisely. However, this also means that the emission velocity is the least well-determined parameter.  A significant increase in velocities will spread the dust out over a larger area of the image.  For fixed values of dust material the high velocities will also create a fainter surface brightness.  The model allows us to determine a velocity at different times throughout the activity.  The velocity can usually be assumed to remain constant so we did not vary the velocity over time in any of our modeling runs.

\subsubsection{Dust Production Rate}

The dust production rate will impact the brightness of the tail and affects its shape the most.  This also means that it is the best constrained parameter.  The function for the dust production rate usually drops off with increasing heliocentric distance.  However, in small outbursts it is possible that the outburst occurs at some point in time and then decreases over the next several days even if the comet is inbound.  In Section \ref{sect:goodness} it is discussed how this parameter can be refined.  

\subsection{49P/Arend-Rigaux as a Candidate for Modeling}

Despite 49P being a well known low-activity comet, it has displayed dust emission features during the 1992, 2004/2005, and 2011/2012 apparitions. The 1992 apparition is convenient for dust modeling because there were 3 observations over a five month period.  In the first image taken in January 1992, the Earth was near the comet's orbital plane and provided constraints on the emission velocity.  During later observations in March, the comet was more face-on allowing better constraints for the dust parameters.  The final image in June showed no dust which limited the number of large grains. The 2004/2005 apparition is good for modeling because it clearly shows two separate components in the tail making the determination of the onset and duration of activity very precise.  Additionally, the 2011/2012 apparition is advantageous for FP modeling because there is a sequence of four observations showing the tail, with two of those images also displaying the jet. Because the morphology of these features can be seen throughout the different images there is a good opportunity to determine the onset of dust production and the grain size distribution.  With three different apparitions modeled we can compare the dust properties and production rates over time. 

Fortunately, 49P has never been observed to have a plasma tail.  This is beneficial for modeling because it reduces the possibility that electromagnetic forces of small charged dust grains moving through a plasma tail will be perturbed enough to be comparable to the radiation pressure forces \citep{Horanyi1985,Ellis1991}.  The lack of a plasma tail also lessens the amount of surface brightness contamination that could come from plasma.

\subsection{Running the Models \label{RunningModels}}

To set up the models we use the geometry of the comet at the time of observation and the image scale and create an input file with a range of $\beta$ and $\tau$ values where $\tau$ represents the onset and offset time of the activity in units of days before or after perihelion.  We can then vary the grain size distribution, the dust production rate, and the velocity distribution with respect to $\beta$ and $\tau$.  Modifying these parameters allows us to find a combination that closely matches the shape of the dust emission and we can determine the goodness of the fit (Section \ref{sect:goodness}). The model steps through the $\beta$ and $\tau$ values coarsely at first, but as the model improves, we can decrease the step size in order to obtain a smooth fit.

\subsection{Goodness of Model Fit}
\subsubsection{Creating Composite Images}

To allow for a comparison between the observed image and the model image without nearby star contamination, we created composite images for each night of observation by shifting to center the comet on the same pixel and median combining the images.   This removes background stars if the star trails do not overlap.  This is important to isolate the comet coma and measure the coma surface brightness and structure without interference of background stars. If the imaging spacing was not large enough we median combined every other frame or every third frame and stacked the results so that the stars were removed.  In some cases we did not have enough observations to do this or the comet was moving slowly, so star residuals still appear in some of the final images. For very bright stars in the image it is difficult to remove them completely, however we did not have any cases where a nearby bright star prohibited us from modeling the dust emission features.  Image dithering has the advantage that the target is not always landing on the same pixels in the chance there are dead pixels.  After stacking, images were trimmed to contain only the comet and the dust features. For modeling purposes we also binned the images into 2$\times$2 bins to reduce modeling time while producing initial models. Once the models started to fit the binned image we used the un-binned image size to obtain the final result. The results of these composite images are represented in comparison to our model results in Figure \ref{fig:FPmodels}. In Figure \ref{fig:jan1992} we do not show the stacked images because there was only one measurement made on January 5, 1992 and two measurements on January 6, 1992.

\subsubsection{Model Residuals \label{sect:goodness}}
We determine an instrument-dependent flux scaling factor by integrating the flux counts over the length of the structure in the dust tail for both the observation and the model.  Then we divide the sum from the observation by the sum from the model.  Taking that factor we scale each pixel in the model to match the observation.  To simulate the noise in the observation we add artificial noise using the IRAF routine, {\sl mknoise} using the read noise and gain to match the observation.  We also simulate an  artificial comet nucleus in the same location in the image as is seen in the observation using the IRAF routine, {\sl mkobject}. The PSF of the artificial comet does not necessarily match the observed comet PSF, it is merely for the viewer to have an easier comparison by eye.  This means that when analyzing the residuals we do not expect the comet to subtract out as well as the dust emission.  

 In order to ensure that the model represents the observation, we subtract the model from the observation and measure the residuals.  We make a cut of the image to only analyze the residuals where there is emission (so any stellar artifacts and the nucleus do not contaminate the residuals) Once the residuals within this cut is within 3$\sigma$ of the background noise, we accept the model as a good fit.  This means that the distributions in the model are only measured to enough precision to understand basic trends.

\subsection{Model Results}

\begin{figure*}[h!tb]
\centering
\includegraphics[width=1.0\textwidth]{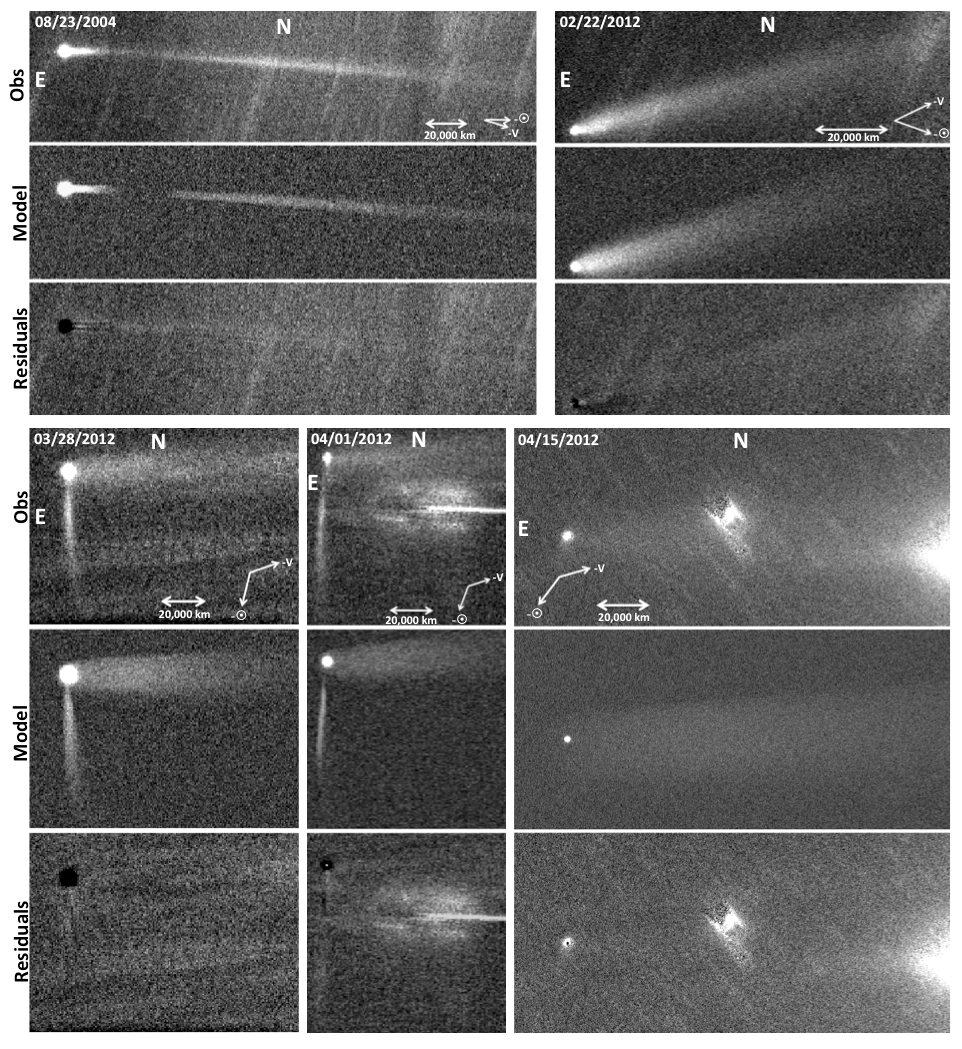}
\caption{Each set of three images shows the observation, FP model, and residuals for a given date from top to bottom.  Dates, orientations, scale (in km), and the negative of the heliocentric velocity (-v) and the extended Sun-target vector or anti-solar direction (-$\odot$) are shown. }
\label{fig:FPmodels}
\end{figure*}

We performed FP Models for five observations taken in 2004 and 2012.  A summary of the results is presented in Table \ref{tab:FPResult} along with FP model results from 1992 for comparison (\citet{farnham1996}).  We note in the table what feature was modeled (tail or outburst/jet).  The images for each observation, FP model, and residual are shown in Figure \ref{fig:FPmodels}.  The observations are the composite images with background stars subtracted.  Streaks in the image are due to residuals from nearby stars, in the April 1 and April 15 observations 49P was near a bright star that did not subtract out well.  The model shows the artificial noise and comet nucleus for comparison to the observation and scaled to the same brightness.  The residuals display the model subtracted from the observation. 

\begin{figure*}[h!tb]
\centering
\includegraphics[width=0.75\textwidth]{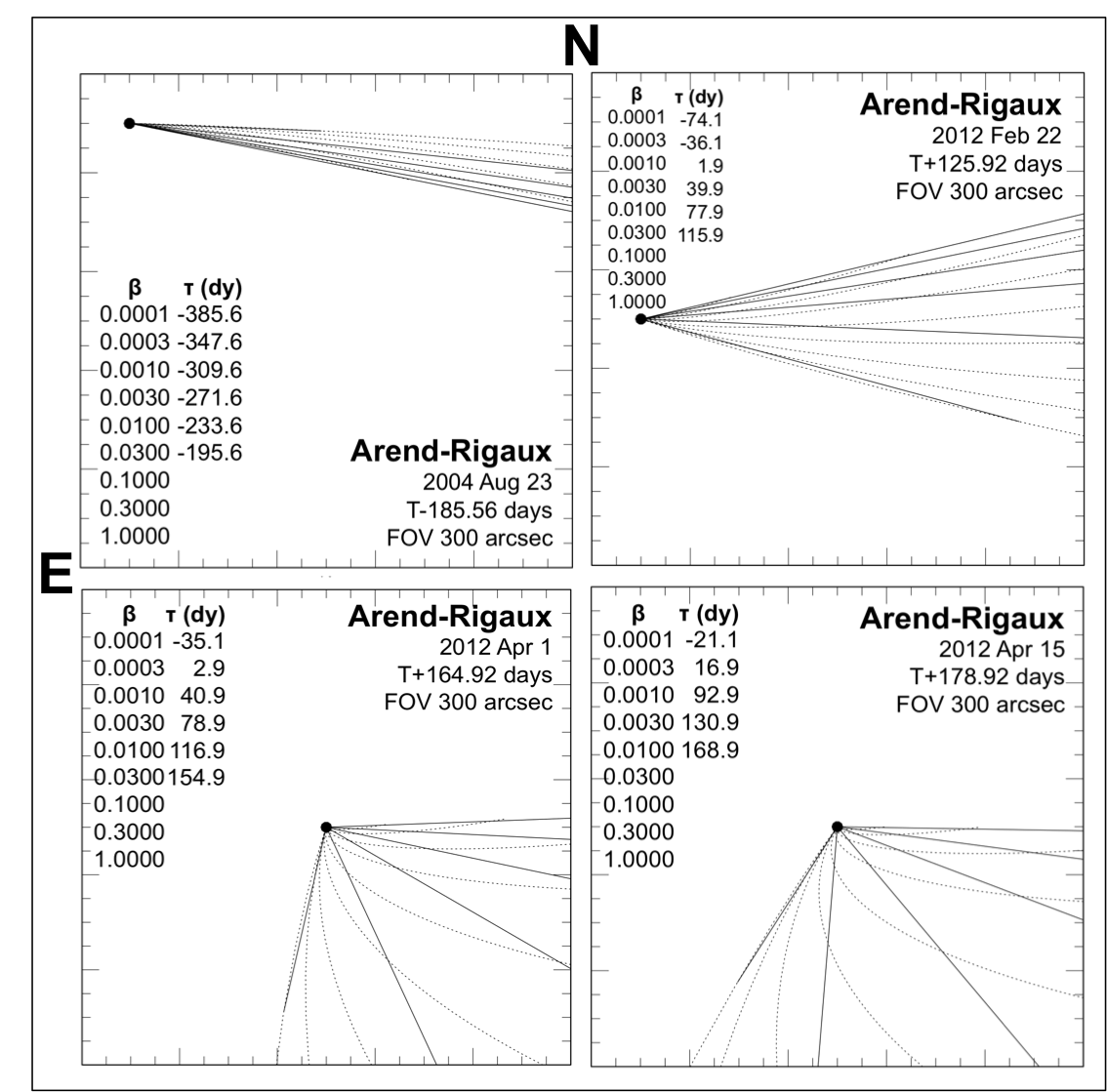}
\caption{Syncurve map of 49P/Arend-Rigaux for the dates modeled in Figure \ref{fig:FPmodels} where April 1, 2012 can be applied to the March 28, 2012 model as well since they are only separated by three days.  The dashed lines represent the syndynes ($\beta$ values) and the solid lines represent synchrones ($\tau$ values). $\beta$ and $\tau$ values are given for each line moving from top to bottom and right to left, clockwise.}
\label{fig:syncurve}
\end{figure*}

For these models, we were able to constrain initial values for $\beta$ and $\tau$ by using syncurve plots of 49P. Syncurves are plots made up of the loci of a suite of particles that are the same size released at different times (syndynes, $\tau$ values) and a distribution of particle sizes released at the same time (synchrones, $\beta$ values).  Using other information such as the orbital geometry of the comet, the superposition of these two sets will map out the possible positions of the tail. A comparison of the observed image to the map of the possible dust loci can provide initial guesses for the model. These curves for the 2004 and 2012 apparition are plotted in Figure~\ref{fig:syncurve}.

\input{tab-FPResults}

\subsubsection{2004 Apparition}

The feature observed in 2004 appears very narrow and interestingly shows up in the observations on August 23, 2004 at 186 days before perihelion.  It appears to have had two outbursts of activity --- there is a gap in brightness at about 10,000 km from the nucleus to about 60,000 km.  Because of this we modeled each event separately with the FP models and combined them into one image.  In both events we see that the grain sizes are small, $\sim$1-8 $\micron$ ($\beta$=0.05-0.38).  The first burst of activity was between 200 and 211 days prior to perihelion, then there appeared to be less activity for about 7 days and then the second burst began at 193 days before perihelion.  Since the observations were taken at 186 days prior to perihelion, it is uncertain how long the second outburst lasted.  The velocities of the grains were low, for the first outburst they were  $\sim$1-2 m s$^{-1}$ and for the second more recent outburst they were $\sim$3-4 m s$^{-1}$ (with lower velocities for larger grain sizes).  We also used a particle size distribution with the functional form given in Equation 3 with $n=2.5$ for the first outburst, and $n=3.8$ for the second outburst for all values of $\beta$. The higher $n$ value signifies that the particles were predominantly smaller dust grains.  There is also a possibility that the gap in activity is due to some large scale variations in the background flux that was not removed during flat-fielding. If this is the case, then the seven days of lower activity could mean that the outburst was actually one continuous outburst.  However since in both outbursts we had similar grain sizes and grain velocities, the results in Table \ref{tab:FPResult} would not change.

\subsubsection{2012 Apparition - Tail}

We model the images from February 22, 2012; March 28, 2012; April 1, 2012; and April 15, 2012 where the comet displays a broad tail feature. These correspond to 126, 161, 165, and 179 days post-perihelion, respectively.  If the activity producing the tail is a gradual process, then the variation between these four dates should not be significantly different.  This meant we were able to constrain our model fits by ensuring that the parameters were very similar for all four observations.  As the syncurve plot implied, we found that the tail consists of large grains $\sim$780 $\micron$ to 4 mm ($\beta$=0.0001-0.0005).  The model shows that the onset of activity began about 100 days before perihelion and continued to produce dust at least 90 days after perihelion.  It is difficult to tell whether the comet was still producing grains later than 90 days because they would still be close to the nucleus if they were large grains.  However, we do not observe any dust emission feature on June 6, 2012, thus the activity most likely decreased significantly around the 90 day mark, making the total duration of activity about 200 days.  The peak of activity was constrained to be between 25 days prior to perihelion.  Since the grains were relatively large particles, the velocities were very low between 0.5 m s$^{-1}$ and 2 m s$^{-1}$.  The particle size distribution followed the distribution in Equation 3 with $n=3.5-4.5$ meaning there were not as many of the large millimeter size grains as there were the smaller grains.    

\subsubsection{2012 Apparition - Jet}

The jet-like feature emanating from 49P in the southerly direction is quite different from the tail.  This feature was not seen during our observations on February 22, 2012 (126 days post-perihelion) and observations from amateur archives do not show the feature until 155 days after perihelion. In our observations we detect this feature easily on April 1 (165 days post-perihelion) but do not see it on April 15 (179 days  post-perihelion).  This implies that this event was relatively short-lived and for it to be almost as long as the tail ($\sim$9.3 $\times$ 10$^4$ km), the grains must either be small or the velocities have to be high.

We model the jet feature using the March 28 and April 1 observations, though the March 28 image is slightly better for modeling due to a bright nearby star that was difficult to subtract out on April 1.  The two models, though close in time, still provide a good way to constrain the model parameters. Running the model, we find that the grains are much smaller than in the tail, with grains ranging from 1.5 to 6.5 $\micron$.  From Equation 3, the particle size distribution for the jet had $n=4.0$. We constrain the peak activity to occur between 143 days and 146 days post-perihelion with a duration of 34 days, starting at $\sim$129 days post-perihelion.  This means that the activity had already ceased by the time we made our observations in March.  In order for this to be the case, the velocity of the grains had to be low for the particles to stay near the nucleus before traveling outward.  We determined grain velocities of only about 1-4 m s$^{-1}$.  

 We note that in all of the activity the dust velocities are quite low.  Other active objects at similar distances display slow dust velocities such as P/2016 J1 (PANSTARRS) \citep{Hui2017}, 133P/(7968) Elst-Pizarro \citep{Jewitt2014}, and 313P/La Sagra \citep{Jewitt2015} - all with dust velocities of $\sim$1.7 - 2.5 m s$^{-1}$. For P/2016 J1 (PANSTARRS) and 313P/La Sagra we recognize that the grain sizes are larger than the grains in the 2012 jet feature so the comparison is not exact but exemplifies how we can have slow moving grains.  Further, measurements from the Rosetta mission of 67P/Churyumov-Gerasimenko had dust velocities between $\sim$2-10 m s$^{-1}$ \citep{Rotundi2015}.  Faster grains would leave the field-of-view more quickly and thus we are not as sensitive to them.

\section{Ice Sublimation Models\label{sect:icesub}}

Comet 49P has reportedly been known as a low-activity comet and \citet{lowry2003} suggested follow-up work to probe whether it is making a transition to an asteroidal state or if the activity is remaining constant.  Using a surface ice sublimation model \citep{meech1986,meech2004,meech2017} we can test whether the fractional active area decreases over time and what ice drives the activity.  This model has many free parameters, which when constrained can provide us with a basic understanding of what volatile species drives the activity. The lead driver of activity for comets close to the sun is sublimation of ices where the ice transitions between a solid and vapor state. When sublimation occurs either from the surface or near-subsurface layers, the gas escapes and pushes dust off with it.  The dust escapes the surface gravity of the comet and appears in the coma and tail. Assuming no contribution from gas fluorescence, the brightness of the comet comes from the sunlight reflected off of the nucleus and scattered light from the dust.  This total coma brightness,$ m_{\rm coma}$ can be expressed in the following equation as a function of mass loss \citep{meech1986}:

\begin{equation}
m_{\rm coma}=C-2.5\textrm{log}_{10}\left[\frac{p_\lambda(dM/dt)t}{ D_gar^2\Delta^2}\right].
\end{equation}

\noindent  The constant, $C$ depends on the apparent magnitude of the sun in the filter we use, among other terms.  For the Kron-Cousins R-band, the constant is 30.3, if a different filter is used then this constant is adjusted accordingly.  The value $t$ represents crossing time and is the aperture size divided by the grain velocity.  We assume the Bobrovnikoff approximation \citep{Bobrovnikoff1954} for velocity, $v\approx r^{-0.5}$ km s$^{-1}$ (for $r$ in au).  The other parameters $p_\lambda$, $D_g$, $a$, and $\Delta$ correspond to the albedo, grain density, grain radius, and geometric distance respectively.  The mass loss rate is defined from the energy balance equation for ice sublimating off of the surface of a nucleus in thermal equilibrium, ignoring heat conduction: 

\begin{equation}
\label{equation:sub}
F_\odot(1-A)/r^2=\chi[\epsilon \sigma T^4+L(T)(dm_s/dt)]
\end{equation}

\noindent 
 The left side of the equation is the absorbed solar flux where $F_\odot$ represents the solar flux constant, and A is the Bond albedo (which for the modeling is $qp_\lambda$, where $q$ is the phase integral and has been measured for low albedo asteroids to be $\sim$0.4 \citep{Shevchenko2019}).  The right side of the equation represents the black-body energy and the energy going into sublimation.  The parameter $\chi$ expresses how fast or slow the nucleus rotates.  If it is slow then heat will only be deposited on one face of the nucleus ($\chi=2$) but if it is fast then heat will be evenly distributed ($\chi=4$).  The infrared emissivity is given by $\epsilon$ and the Stefan-Boltzmann constant is given by $\sigma$.  L(T) represents the latent heat as a function of temperature for common ices \citep{meech1986}.  Finally the parameter $dm_s/dt$ provides the mass loss rate per unit area and relates to the sublimation vapor pressure and thermal gas velocity. This model is not intended to explore the depth of ices, but rather to model which ices contribute to the overall activity in the comet.

The equations above allow for several free parameters in the model including properties of the nucleus (ice type, albedo, emissivity, phase integral, density, and radius) and the dust grains (density and radius). We also have the fractional active area of the nucleus as a free parameter.  Fortunately most of these values have been directly measured for 49P and other values can be assumed based on our previous knowledge of comets. The two main parameters that we adjust are the fractional active area and the grain size. The other free parameters and our accepted values that we use are provided in Table \ref{tab:sublimation} with references.   It is important to note that the density from \citet{Thomas2013} was measured for 9P/Tempel 1 and the coma phase integral from \citet{Meech1987} was derived from 1P/Halley.  

\input{tab-sublimation}

To determine an approximate grain size for 49P we use the results from our FP models. We found that the outbursts in 2004 and 2012 consisted of 1-8 $\micron$ size grains and \citet{millis1988} found a consistent range of 3-5 $\micron$ size grains.  However the tail feature in 1992 and 2012 have grain sizes greater than 40 $\micron$ but are still dominated by smaller grains so we vary the grain sizes in the models from 3 to 8 $\micron$ to find the best fit.  The fractional active area and grain sizes are related. As the grain sizes increase then a larger gas flux is needed to lift the grains from the surface. Additionally, since the surface area to mass ratio depends inversely on grain size, the number of grains has to increase for the massive grains to produce the same amount of scattered light (which means a higher gas flux, and higher fractional surface area). Therefore we can still determine whether the fractional active area decreases over time but knowing the precise grain sizes allows us to more precisely determine how much of the comet is active. 

 The  typical ices we would expect to sublimate from comets are CO, CO$_2$ and H$_2$O. It is difficult to distinguish between CO and CO$_2$ ices since at this distance from the sun the slopes of the sublimation curves are the same but we can easily determine whether H$_2$O is present. Many space-based observations have found CO$_2$ dominates activity over CO so we only run models with either H$_2$O or CO$_2$. After running the models using a particular ice, grain size, and fractional active area (and including the parameters in Table \ref{tab:sublimation}), the $R$ magnitude throughout the apparition and the gas production rate at incremental heliocentric distances are produced. We plot the $R$ magnitude of  our observations and the $R$ magnitude produced by the model versus the true anomaly (using the same definition of TA as Table \ref{tab:photom}). When the model fits the most data points and generally represents the same trend as our observations we accept it as a fit. In most apparitions we have too few data points to do a $\chi^2$ minimization so our fit is designed only to represent a general range of values for the fractional active area and determine what ice drives the activity on the comet. 

The gas production rate is a useful output because we can match the values to previously observed production rates at a particular heliocentric distance. For example, \citet{ahearn1995} found logQ(H$_2$O)=27.25~molecules~s$^{-1}$ at 1.56 au on August 9, 1991. This allowed us to identify an exact fractional active area to match the gas production rate and find the most dominant grain size for a single apparition. Similarly, \citet{reach2013} found logQ(CO$_2$)=26.13$\pm$0.35~molecules~s$^{-1}$ at 1.69 au on November 29, 2011. This allows us to estimate how much CO$_2$ should be included during this apparition.

\subsection{Data for Modeling}

Our observations combined with literature and amateur data span a wide range of dates from 1984 through 2018. These dates cover six apparitions of 49P for which we have our own data in four of them.  We define a single apparition from aphelion to aphelion and the range of dates with the number of nights of observations are given in Table~\ref{tab:apparition}. We will refer to each apparition by the ID number given in column 1.

\input{tab-apparition}

Some apparitions have several of our own data points that cover a full range of the orbit for a single apparition, while others only have a small coverage. We also have two literature values from the third apparition from \citep{lowry2001, lowry2003} (Table~\ref{tab:photom}).  In order to fill in some of the heliocentric light curve, we use measurements from the Minor Planet Center (MPC) database\footnote{https://tinyurl.com/49P-MPC-Results for all data with a designation of N (nucleus) for the magnitude measurement, and for the range of dates covering the third through sixth apparitions}.  These observations have reported magnitudes with an N (nucleus) or a T (total) to signify whether the magnitude encompasses just the nucleus or if it has a larger aperture including more of the coma or background sky. In order to avoid possible nearby star contamination with larger apertures, we only include the observations of the nucleus.  The spectral filters used for observations are not included in the MPC archives but we expect the brightness in different bands to vary by about 0.5 magnitudes at most, which is acceptable to understand basic trends over time.  In the first two apparitions the MPC data is sparse and very scattered so we do not use the MPC values for these apparitions. In the third apparition there are only a few MPC data points but they are similar to our measured photometry and at a similar true anomaly. For the fourth apparition we do not have any of our own measurements but the MPC data has some coverage of the heliocentric light curve.  For the fifth apparition the MPC data does not perfectly align with our measurements and this is due to unknown filters. To adjust for this factor we shift the MPC magnitudes until it matches our measurements for the fifth apparition, which is 0.5 magnitudes.  With this known shift, we also adjust the fourth apparition measurements by the same scale factor. In the sixth apparition, 49P was mainly a daytime comet and very few observations were possible so we only have a small amount of data from the MPC.   For the fourth and sixth apparition, we use these data to understand the shape of the light curve to tell us what gas is driving the activity but we cannot make claims about the fractional active area using H$_2$O; however, Section~\ref{sect:water model results} discusses how we can constrain the CO$_2$ fractional active area.  There is data dating back before our first apparition but we do not include these points because there are very few, and they have higher uncertainties. 

\subsection{Summary of Model Results \label{sect:water model results}}

The dependence on the geometry between the observer and the target changes the brightness of the comet significantly for different apparitions, therefore we require a separate model for each apparition. We began by modeling our second apparition using H$_2$O ice because we have an H$_2$O production rate measured for August 9, 1991 \citep{ahearn1995}. By iterating through values for the fractional active area we found that the measured H$_2$O production rate aligned with the model with the fractional active area of 0.0063. We then varied the grain size between 3-8 $\micron$ (from the FP model) and found that the best fit grain size is 8 $\micron$. This is a simplification because there is typically a full size distribution of grains but this value represents the most prominent size.  We then modeled the remaining apparitions using H$_2$O ice and found an estimate of the fractional active area for each grain size between 3-8 $\micron$. The best fits are determined by visual inspection but the model starts to change significantly in appearance when the fraction is changed by $\sim$0.0005  ($\sim$0.001 for Apparition 1) so we know that the fits are good to within this error.  

\begin{figure*}[h!tb]
\centering
\includegraphics[width=0.95\textwidth]{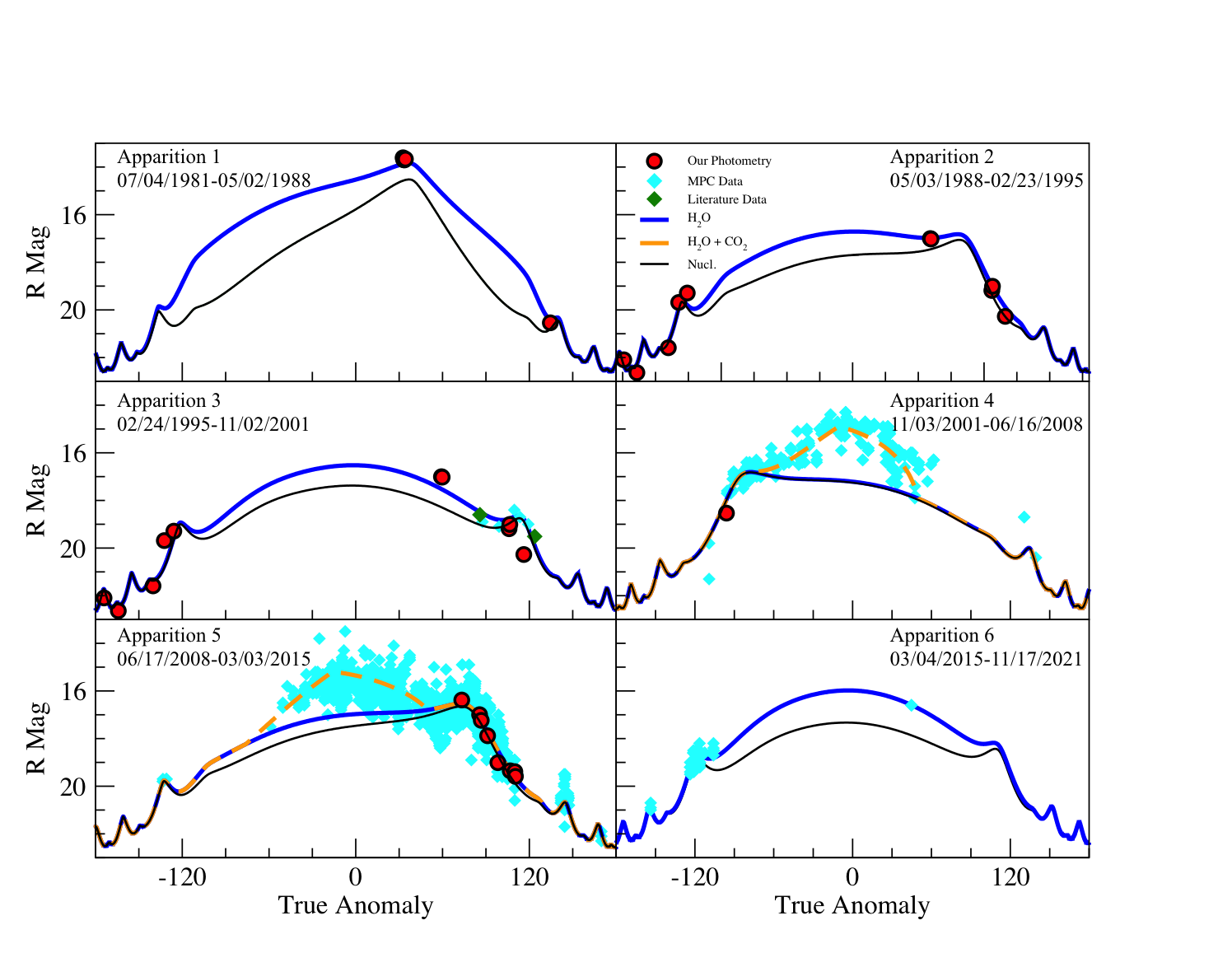}
\caption{49P/Arend-Rigaux photometry for six apparitions compared to ice sublimation models for H$_2$O (blue line) and a combination of H$_2$O + CO$_2$ (orange dashed line).  Measurements reported in this paper for the first time are shown with red circles. Cyan diamonds represent data from the Minor Planetary Catalog, green diamonds are taken from the literature (Supplemental Observations in Table \ref{tab:photom}). The black line represents a bare nucleus.  In Apparition 4 the bare nucleus and H$_2$O model are not distinguishable in the figure. A True Anomaly of 0 represents perihelion.}
\label{fig:allapp}
\end{figure*}

Figure~\ref{fig:allapp} shows the ice sublimation models for each apparition with our corresponding photometric measurements; the last four apparitions include MPC data and literature values. The figure shows the true anomaly versus the R magnitude as a representation of the heliocentric light curve.  For each apparition the H$_2$O model dominates the activity during most of its orbit while the fourth and fifth apparitions show a combination of H$_2$O and CO$_2$ to explain the excess brightness near perihelion. There is not a sufficient amount of data to determine if CO$_2$ is present for other apparitions near perihelion.  We see from the MPC data that the light curve shape does not match the H$_2$O ice trend between a true anomaly of $\sim$-60~to~$\sim$60.  

Therefore, we ran our models for the fourth and fifth apparition using a delayed onset of CO$_2$ ice to combine with H$_2$O ice, similar to previous works (e.g. \citet{meech2013}, \citet{Snodgrass2013}, \citet{meech2017}). CO$_2$ ice is more volatile than H$_2$O and would sublimate beyond 10 au if it is on the surface but if it were buried in a deeper layer, the model would only show CO$_2$ outgassing later in the orbit. If the comet has an insulating porous dust layer or even a water ice layer (as seen during the Rosetta mission \citep{Biele2015,Gulkis2015}) then it would take time for the heat to penetrate deep enough to sublimate the CO$_2$.  Equation \ref{equation:sub} does not incorporate heat conduction into the interior of the comet but we are not attempting to determine the depth of ice or the time it takes for the ice to reach a buried layer, but we know that an additional ice component is necessary in order to model the increase in brightness near perihelion.  Because the model deviates away from a pure H$_2$O model at TA$\sim$-60 we allow for the CO$_2$ ice to begin sublimating exponentially near TA~=~-60 as heat starts to penetrate the deeper ice layers and reaches its peak CO$_2$ outgassing near perihelion. Then we let the outgassing decay exponentially to TA~=~+60. 

When we apply this adjustment and add both the CO$_2$ and H$_2$O ice together in the model we match the data very well. From \citet{reach2013} we use the CO$_2$ gas production rate from the fifth apparition, log Q(CO$_2)$=26.13, to find the fractional active area (similar to determining the fractional active area using the H$_2$O gas production rate in the second apparition). We found that the fractional active area is 0.00023 -- an order of magnitude less than the fractional active area of H$_2$O in the third and fifth apparitions, meaning there is significantly less contribution from CO$_2$ than water.  We do not know whether this low activity from CO$_2$ is due to a small presence of the ice, or if there was not enough heat to penetrate into deeper layers of CO$_2$. We cannot distinguish whether the ice is CO or CO$_2$ but it is more likely CO$_2$ based on recent space missions \citep{Ahearn2012} and the gas production rate from \citet{reach2013}. 

We consider that the first apparition might display the presence of CO$_2$ near perihelion. To match the observation on March 7, 1986 we require H$_2$O for a good fit but the H$_2$O model does not produce bright enough results to match the data from January, 1985 (\citet{jewitt1985}). These observations were closer to perihelion so it is possible that CO$_2$ drove the activity as in the fourth and fifth apparition.  We tested this apparition by modeling a smaller fractional active area of H$_2$O and instead exponentially increasing the CO$_2$ near perihelion. The only way the 1985 data fits the model is with a fractional active area of CO$_2$~$\sim$0.008, over an order of magnitude greater than the fractional active area of CO$_2$ in the later apparitions.  Unfortunately with such small temporal sampling we cannot conclusively determine whether CO$_2$ has a role in driving the activity during this apparition.  It is however clear that the fractional active area is still greater than the later apparitions whether we use a pure H$_2$O model or a H$_2$O and CO$_2$ combination model.

\begin{figure}[h!tb]
\centering
\includegraphics[width=0.45\textwidth]{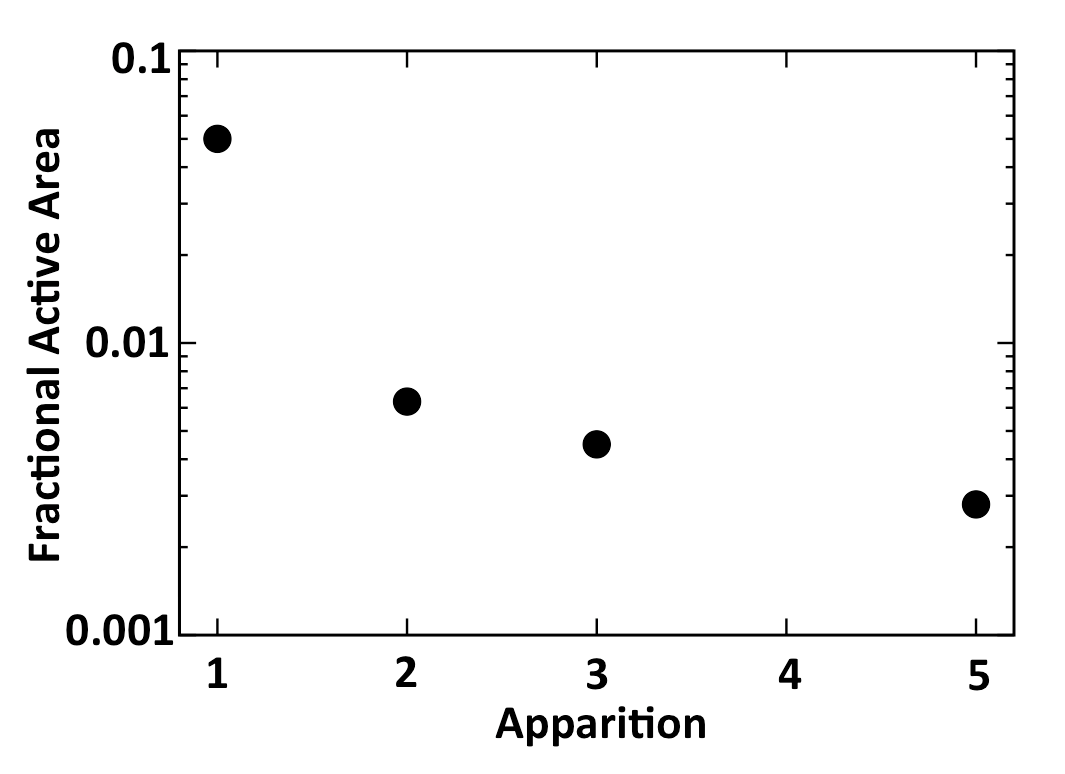}
\caption{This represents the decay of the fractional active area over time.  We show the fractional active area from Table~\ref{tab:fracactive} for a grain size of 8 $\micron$. }
\label{fig:fracactive}
\end{figure}

Knowing the fractional active area of each apparition allows us to identify trends to determine if the comet has become more or less active over time. The results of our models with the grain size and fractional active area or H$_2$O is given in Table~\ref{tab:fracactive}. Based on these values we also plot the apparition versus the fractional active area using the first, second, third, and fifth apparitions (Figure~\ref{fig:fracactive}).  Though the fourth and sixth apparitions are important to understand the ices that are sublimating, We do not include these two apparitions in our figure.  This is due to a lack of data since we rely on the MPC data almost entirely which could be offset by an unknown scaling factor.  They also have no constraints from measured gas production, so we can only glean information from the shape of the curve to definitively say what ices are most likely involved but it is not appropriate to use them to determine the fractional active area.  Thus the four apparitions shown with properly calibrated data demonstrates that each subsequent apparition displays a decrease in the fractional active area.  This is discussed further in Section \ref{demise}.

\input{tab-fracactive}

\section{Discussion}

\subsection{Finson-Probstein Model}

\subsubsection{Comparison to Previous 49P Models}

\begin{figure}[h!tb]
\centering
\includegraphics[width=0.45\textwidth]{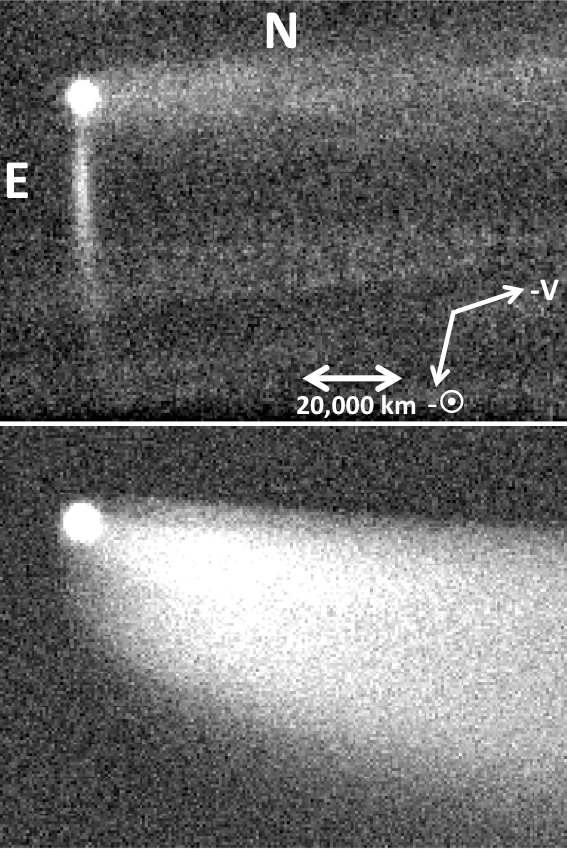}
\caption{The top shows the observation from March 28, 2012.  The bottom is the FP model using the same grain sizes as the 1992 apparition (40-400~$\micron$) and scaled to match the brightness of the observation. }
\label{fig:comparison}
\end{figure}

Using the FP models from \citet{farnham1996} for the 1992 apparition, and the 2004 and 2012 models from this work, we can notice some similarities and differences in the outgassing events.  In 1992 the majority of the dust production took place near perihelion then mostly ceased, producing only small amounts of dust up to $\sim$157 days after perihelion.  Similarly in 2012 the dust production for the tail peaked near perihelion and had a long duration of slow outgassing.  In both cases the grain sizes were considerably large (40~$\micron$-400~$\micron$ in 1992, 780~$\micron$-4~mm in 2012), but the grains from 2012 were distinctly larger.  To show that the 2012 tail must have larger grains, we simulated what the tail would look like if we used the same FP parameters of $\beta$ and $\tau$ as the 1992 apparition (Figure \ref{fig:comparison}).   It is clear these parameters are not a good representation of the tail from 2012 and that even larger grains are necessary to have the tail pointing more narrowly in the westward direction.

 In 1992 there was an additional distinct narrow feature along the northern edge of the tail that was found to be composed of smaller grains than the rest of the tail.  This matches closely the features seen in both 2004 and the jet in 2012.  In each of these cases the grain sizes ranged from $\sim$1-8 $\micron$ and were short outburst events.   These outbursts are perhaps caused by the presence of small pockets of ice deep beneath the surface that have a later onset due to the time it takes for heat to penetrate deep enough.  This mechanism does not explain the 2004 event since the feature is seen over 200 days before perihelion.  It is possible that it is caused by a deeper reservoir of CO$_2$ ice since that will sublimate sooner than H$_2$O.  We know from the ice sublimation models that CO$_2$ must be present on 49P, so this is a reasonable explanation.  The difference between these two types of features on 49P implies that the ice reservoir could be different for each event --- the tail features with large grains most likely come from shallow subsurface ice, while the jets are produced from small, localized regions deeper beneath the surface.

\subsubsection{ Heliocentric Distances for which Activity is Observed}

In some apparitions 49P had visible activity prior to perihelion, while other apparitions the activity is dominant post-perihelion.  This leads us to determine whether the heliocentric distance plays a role in the activity.  The heliocentric distance for the activity in 1992 is closer than the distance in 2012 ($r$=1.7 au versus $r$=2.32 au, both post perihelion).  However, from the observations in 1985 there is no visible activity even at a heliocentric distance of $\sim$1.5 au \citep{jewitt1985}.  When modeling the 1985 data with ice sublimation models the comet appears brighter than the H$_2$O model predicts meaning it possibly displayed some low activity at the time of observation, but is inconclusive.  Adding data from \citet{reach2007,reach2013} 49P shows activity pre-perihelion up to $\sim$1.7 au and in 2004 the activity occurs pre-perihelion at a distance of 2.35 au. This made for a very well sampled data set constraining the previously seen activity to occur approximately between 200 days prior to perihelion and 200 days after.  Figure \ref{fig:orbit} marks the observations that showed activity in relation to their orbital position.  The distances at which 49P is active are relatively low compared to other high activity comets such as 9P/Tempel 1, 1P/Halley, and 103P/Hartley 2 which show activity for 1992, 1069, 1484 days respectively surrounding perihelion. However 49P does not have the shortest activity: comet 2P/Encke is usually only active for about 180 days despite having a closer perihelion.  Therefore 49P has a rather average duration of activity. 

\subsubsection{Outburst Duration and Mechanisms}

Comets can display outbursts with varying durations but it is still unclear what mechanisms drive the activity.  It is likely that the jet and tail are caused by dust dragged out by sublimating gas, but we do not know if the sublimation is due to a quasi-equilibrium between energy input and latent heat or a pressure build-up causing an outburst.  The {\it Deep Impact} mission showed many small outbursts on 9P/Tempel 1 lasting on the order of seconds to minutes \citep{ahearn2005}. Water vapor from ices sublimating from the near surface region in the late afternoon would reach the surface after it had moved out of sunlight and then re-freeze on top of the dust \citep{prialnik2008}.  This resulted in short outbursts as the ices sublimated when the terminator rotated into sunlight--observed both during the 9P/Tempel 1 and 67P/Churyumov-Gerasimenko encounters \citep{Deshapriya2018, Fornasier2016}. Other mechanisms have also been proposed to explain short-lived outbursts but these outbursts are in general only detectable from space-based observations so these are not comparable to the type of outburst we observed on 49P. In general most observed outbursts, where the brightness of the comet increases significantly (sometimes several magnitudes), will appear to last on the order of tens of days after a short pulse of extra mass loss \citep{hughes_1991}.  This duration is more similar to what we find for 49P.

Some comets have frequent activity with several outbursts a year like 29P/Schwassmann-Wachmann 1 \citep{trigo2008, kossacki2013} while others display very few outbursts (e.g. 2P/Encke, \citet{levison2006}).  Outbursts of 67P/Churyumov-Gerasimenko were discussed by \citet{Knollenberg2016} and \citet{Vincent2016} but an unambiguous mechanism could not be identified.  The distances at which outbursts occur also vary and the different comets can provide clues as to what drives activity.  For comets near the sun, outbursts could be easily explained by pressure buildup underneath the surface and as heat penetrates deeper the pressure releases the volatile gas in a sudden outburst.  At distances where H$_2$O does not sublimate then the ice may consist of CO or CO$_2$ ice \citep{schleicher2009}. There are many other mechanisms that could describe cometary outbursts such as polymerization of or Hydrogen Cyanide (HCN), thermal stresses, annealing of amorphous water ice, meteoritic impacts, and sinkhole collapse \citep{gronkowski2010,ivanova2011,Vincent2015}. With all of these various mechanisms it is difficult to determine the exact cause of the outbursts on 49P.  If the material on the comet is highly porous then pressure build-up is implausible. We do however know that 49P is CO$_2$ rich so the release of this volatile could be at least one of the mechanisms for these short outbursts \citep{reach2007,reach2013}.


\subsection{Ice Sublimation Model}

\subsubsection{\label{demise}Demise of Comets}

We benefit from having data spanning several apparitions of 49P including MPC data to allow us to constrain our ice sublimation models. Because we use the FP models to determine the dust grain sizes we can determine a fractional active area and find any trends over time. Figure~\ref{fig:fracactive} demonstrates that each apparition has a decrease in the fractional active area implying that the comet may be  potentially running out of volatiles. The other possibility is that 49P is merely becoming dormant and has a rubble mantle with deep ice layers or pockets that could eventually become active through mantle cracking.  After the resealing of the mantle, the comet becomes dormant again \citep{jewitt2004}.  If the comet decreases its perihelion distance from gravitational perturbers (such as close approaches with Jupiter) and receives more heat from the sun, the mantle can crack and reseal. Over time this will deplete the available ices at a faster rate than if it remains at the same heliocentric distance.  Over the past six apparitions 49P has had a perihelion distance decrease of $\sim$0.02 au which increases the amount of flux it receives by $\sim$3\% making this mantle cracking and resealing method a possibility, though usually a larger decrease would be necessary. If this mantle cracking is true then 49P may in fact be running out of volatiles which is why the fractional active area has been decreasing. If this mechanism does not play a role then 49P could still be losing its volatiles just more slowly over time.

Comets are thought to be active between 3000 and 30,000 years and the best estimate of the ratio of extinct to active Jupiter Family Comets is 3.5 \citep{levison1997}. So if the majority of comets are actually dead then our results of a comet with decreasing activity are vital to understanding the process of their demise.

There are other comets that may also be part of this class of dying or dormant comets, for example, 107P/Wilson-Harrington had an outburst during its discovery in 1949 but since then has shown no further activity. Also the object,  28P/Neujmin 1 has shown only small amounts of activity, which was recorded in 1984 \citep{Ahearn1984}. It is possible that 49P could soon appear like these two objects.

We must however consider that we witnessed outgassing in 1992, 2004, and 2012, so it is not dormant yet but since the active areas are decreasing, the volatiles may soon run out. 

 We note that we also do not have reliable data from the fourth and sixth apparitions which may counteract this trend if it were able to be constrained.  The first apparition also limits us where the errors on the fractional active area are larger due to the significant dependence on the data from 1985.  Therefore, while the decay is certainly intriguing, we are cautious to conclude that the activity is indeed decreasing and continuous monitoring will help confirm this trend.    

\subsubsection{Presence of Carbon Dioxide \label{CO2}}

\citet{reach2013} found that 49P is a CO$_2$-rich comet.  They observed the comet in 2004, the same apparition that \citet{stevenson2007} witnessed a narrow jet-like feature. Our ice sublimation models also indicate that CO$_2$ is responsible for extra activity near perihelion, due to a delayed onset of sublimation resulting from the ice coming from deeper below the surface.  CO$_2$ could therefore explain what caused the outburst of the jet in 2004 and similarly lead us to conclude that the jet in March, 2012 was also a product of CO$_2$ outgassing. This may explain the early onset of activity before perihelion in 2004 since the CO$_2$ is more volatile.  However, CO$_2$ appears not to be constant over the entire apparition. It is possible that the CO$_2$ ice is deep within the nucleus and takes time to heat up and then it diminishes after perihelion \citet{meech2011}. This would explain why the shape of the ice sublimation model can be fit using H$_2$O, except near perihelion where CO$_2$ must be included. 

\section{Conclusion}

Comet 49P/Arend-Rigaux has been known for over sixty years and despite its classification as a low-activity comet, has shown outgassing events in several of the more recent apparitions.  Using Finson-Probstein dust dynamical modeling we found that the tails seen in 1992 and in 2012 are similarly produced by larger grains (40-4000 $\micron$) with a peak activity around perihelion and are slow ongoing events.  In 2004 a more jet-like outburst occurred, which similarly matched a jet feature in 2012.  These two events were dominated by smaller grains (1-8 $\micron$) with short duration on the order of about a month.  These short outbursts are comparable to other comet outbursts and are potentially caused by heat reaching deeper layers of CO$_2$ ice.  When we apply ice sublimation models we determine that CO$_2$ does play a role in driving some activity on 49P so it could have contributed to the jets.  \citet{reach2013} find that there is an abundance of CO$_2$ so the low activity on 49P could be a result of having deep ice layers that usually take a significant amount of heat to be activated.  If the heat does not penetrate deep enough then we would only observe activity from a thin layer of H$_2$O ice which would match 49P's reputation as a low activity comet.

Using information from the literature, and grain sizes from the Finson-Probstein models, we model the fractional active area on the comet for each apparition produced by H$_2$O sublimation.  In early apparitions, an average of $\sim$3\% of the surface is active. Over the following apparitions average fractional active area decreases down to $\sim$0.2\% of the surface. While this result is still highly model dependent, this decrease in activity may indicate that 49P is depleting its reservoir of shallow ices and may still have deep pockets of ice that outgas from mantle cracking.  It could also mean that we are witnessing the transition of a comet to a more asteroidal state, which will only be revealed with continuous monitoring. 

{\it Acknowledgements} -- The authors would like to thank Jason Chu for helpful discussions and assistance in data analysis.   We also thank the reviewers for their thorough comments. This material is based upon work supported by the National Aeronautics and Space Administration through the NASA Astrobiology Institute under Cooperative Agreement No. NNA04CC08A issued through the Office of Space Science, by NASA Grant Nos. NAGW 5015, NAG5-4495, NNX07A044G, and NNX07AF79G, Image processing in this paper has been performed, in part, using the IRAF software. IRAF is distributed by the National Optical Astronomy Observatories, which is operated by the Association of Universities for Research in Astronomy, Inc. (AURA) under cooperative agreement with the National Science Foundation.   The authors recognize  that  the summit of Maunakea has always held a very significant cultural role for the indigenous Hawaiian community. We are  thankful to have the opportunity to observe from this mountain.

\bibliographystyle{aasjournal}
\bibliography{49P.bib}

\end{document}

%% file: tab-photom_new.tex
 \begin{table*}[h!t]
\tabcolsep=0.06cm
\tabletypesize{\small}
\centering
\caption{Photometry of 49P/Arend-Rigaux}
\label{tab:photom}

\begin{tabular}{cccccccccccccccc}
\tableline
UT Date	  \footnote{$\ast$ Active Obs shown in Figure~\ref{fig:orbit}}                                       &	
Julian Date	                                     &	
r \footnote{Heliocentric Distance (au)}          &	
$\Delta$ \footnote{Geocentric Distance (au)}     &	
$\alpha$ \footnote{Phase angle (deg)}	         &	
TA \footnote{True Anomaly (deg): the angle from the direction of periapsis to the target, -180$^{\circ}$ to 0$^{\circ}$ is pre-perihelion and 0$^{\circ}$ to +180$^{\circ}$ is post-perihelion}	             &
Tel \footnote{MHO=McGraw Hill 1.3m on Kitt Peak, KPNO=KPNO 2.1m on Kitt Peak, UH2.2m=University of \hawaii~2.2m, CTIO1.5=1.5m CTIO4=4m at the Cerro Tololo Inter-American Observatory, Cal=Calar Alto 1.2m telescope, WHT=4.2m William Herschel Telescope, JKT=1m Jacobus Kapteyn Telescope} &
Inst \footnote{CCD Detector type}                & 
Scale \footnote{($\arcsec$ pix$^{-1}$)}          & 
G \footnote{Gain (e$^{-}$ ADU$^{-1}$)}           & 
RN \footnote{Read Noise (e$^-$)}                 &
Sky \footnote{Weather: p=photometric, c=cirrus}  &	
\#  \footnote{Number of images}	                 & 
Exp \footnote{Total exposure time (seconds)}     & 
R Mag \footnote{$\ast$ Johnson R ( $\lambda_{cent}\sim$ 0.658 $\micron$, $\Delta \lambda\sim$ 0.138 $\micron$), + Kron Cousins R ( $\lambda_{cent}\sim$ 0.647 $\micron$, $\Delta \lambda\sim$ 0.125 $\micron$), $\ddagger$ Sloan Digital Sky Survey r  ($\lambda_{cent}\sim$ 0.626 $\micron$, $\Delta \lambda\sim$ 0.134 $\micron$)}                   & 
Ref \footnote{$^{(1)}$ \citet{lowry2001}, $^{(2)}$ \citet{lowry2003}, $^{ (3)}$ \citet{stevenson2007}, $^{(4)}$ \citet{reach2013}, $^{(5)}$ \citet{reach2007}}\\
\tableline
01/18/85  & 2446083.5 & 1.538 & 0.567 & 10.04 & 32.7 &  MHO      &  MASCOT    &  1.600  &  ?  &  12.0  &  p  &  6  &  420  &  13.610 $\pm$ 0.040  $\ast$  &  - \\
01/19/85  & 2446084.5 & 1.541 & 0.569 & 9.32 & 33.3 &  MHO      &  MASCOT    &  1.600  &  ?  &  12.0  &  p  &  14  &  980  &  13.690 $\pm$ 0.040  $\ast$  &  - \\
01/20/85  & 2446085.5 & 1.545 & 0.571 & 8.62 & 34.0 &  MHO      &  MASCOT    &  1.600  &  ?  &  12.0  &  p  &  25  &  1750  &  13.700 $\pm$ 0.040  $\ast$  &  - \\
01/21/85  & 2446086.5 & 1.549 & 0.573 & 7.94 & 34.6 &  MHO      &  MASCOT    &  1.600  &  ?  &  12.0  &  p  &  27  &  1890  &  13.670 $\pm$ 0.040 $\ast$  &  - \\
03/07/86  & 2446497.0 & 3.985 & 3.611 & 13.92 & 134.6 &  KPNO  &  TI \#3    &  0.396  &  4.3  &  5.0  &  c  &  3  &  900  &  20.538 $\pm$ 0.089 +  &  -  \\
09/11/88  & 2447415.8 & 5.713 & 5.468 & 10.02 & -174.1 &  UH2.2  &  GEC        &  0.200  &  1.2  &  6.0  &  p  &  10  &  3000  &  22.092 $\pm$ 0.076 +  &  -  \\
04/06/89  & 2447622.9 & 5.449 & 5.556 & 10.38 & -164.2 &  CTIO1.5  &  TI \#2    &  0.550  &  2.9  &  11.0  &  p  &  13  &  3900  &  22.637 $\pm$ 0.104 $\ast$  &  - \\
04/29/90  & 2448010.8 & 4.265 & 4.413 & 13.19 & -140.1 &  CTIO4  &  Tek 512    &  0.177  &  1.9  &  6.7  &  p  &  20  &  3000  &  21.589 $\pm$ 0.045 $\ast$  &  - \\
07/23/90  & 2448095.9 & 3.866 & 2.918 & 6.28 & -132.4 &  UH2.2  &  GEC        &  0.556  &  1.2  &  6.0  &  p  &  33  &  2970  &  19.687 $\pm$ 0.015 +  &  -  \\
09/22/90  & 2448156.8 & 3.545 & 2.787 & 12.03 & -125.8 &  UH2.2  &  GEC        &  0.556  &  1.2  &  6.0  &  p  &  19  &  2700  &  19.291 $\pm$ 0.018 +  &  - \\
01/05/92$\ast$  & 2448627.1 & 1.763 & 1.283 & 33.16 & 59.5 &  UH2.2  &  Tek 1024   &  0.351  &  3.5  &  10.0  &  p  &  1  &  900  &  17.004 $\pm$ 0.007 +  &  - \\
01/06/92$\ast$  & 2448628.1 & 1.769 & 1.279 & 32.93 & 59.9 &  UH2.2  &  Tek 1024   &  0.351  &  3.5  &  10.0  &  p  &  2  &  1200  &  17.024 $\pm$ 0.009 +  &  - \\
06/04/92  & 2448777.7 & 2.759 & 2.400 & 21.25 & 106.1 &  KPNO  &  Tek \#4    &  0.304  &  2.7  &  3.5  &  p  &  4  &  2460  &  19.186 $\pm$ 0.017 +  &  - \\
06/06/92  & 2448779.8 & 2.772 & 2.438 & 21.24 & 106.5 &  UH2.2  &  Tek 2048   &  0.219  &  1.7  &  6.0  &  p  &  4  &  2400  &  19.007 $\pm$ 0.090 +  &  - \\
08/02/92  & 2448836.8 & 3.133 & 3.495 & 16.47 & 116.3 &  UH2.2  &  Tek 2048   &  0.219  &  1.7  &  6.0  &  p  &  4  &  1200  &  20.272 $\pm$ 0.049 +  &  - \\
06/14/96  & 2450249.1 & 5.113 & 4.272 & 6.99 & -156.9 &  UH2.2  &  Tek 2048   &  0.219  &  1.7  &  6.0  &  p  &  2  &  1800  &  21.397 $\pm$ 0.210 +  &  - \\
11/22/97  & 2450774.8 & 2.676 & 2.623 & 21.45 & -106.7 &  UH2.2  &  Tek 2048   &  0.219  &  1.7  &  6.0  &  p  &  2  &  1200  &  19.184 $\pm$ 0.014 +  &  - \\
08/21/00  & 2451777.8 & 5.131 & 4.992 & 11.36 & 158.7 &  UH2.2  &  Tek 2048   &  0.219  &  1.7  &  6.0  &  p  &  3  &  1800  &  21.618 $\pm$ 0.078 +  &  - \\
02/22/12$\ast$  & 2455979.6 & 1.950 & 1.050 & 16.49 & 73.7 &  Cal      &  DLR-MKIII\footnote{e2v CCD231-84 Detector}  &  0.314  &  2.8  &  4.2  &  c  &  16  &  4800  &  16.380 $\pm$ 0.001 $\ast$  &  - \\
03/28/12$\ast$  & 2456014.5 & 2.185 & 1.290 & 15.14 & 85.9 &  Cal      &  DLR-MKIII  &  0.314  &  2.8  &  4.2  &  c  &  12  &  3600  &  16.996 $\pm$ 0.003 $\ast$  &  - \\
04/01/12$\ast$  & 2456018.5 & 2.212 & 1.334 & 15.94 & 87.1 &  Cal      &  DLR-MKIII  &  0.314  &  2.8  &  4.2  &  c  &  18  &  5400  &  17.235 $\pm$ 0.003 $\ast$  &  -\\
04/15/12$\ast$  & 2456032.9 & 2.311 & 1.515 & 18.80 & 91.3 &  UH2.2  &  Tek/WFGS2  &  0.219  &  1.7  &  6.0  &  p  &  6  &  2700  &  17.885 $\pm$ 0.003 $\ddagger$  &  -  \\
06/05/12  & 2456084.4 & 2.659 & 2.366 & 22.33 & 103.8 &  Cal      &  DLR-MKIII  &  0.314  &  2.8  &  4.2  &  c  &  15  &  4500  &  19.010 $\pm$ 0.018 $\ast$  &  - \\
06/21/12  & 2456100.4 & 2.764 & 2.662 & 21.48 & 107.0 &  Cal      &  DLR-MKIII  &  0.314  &  2.8  &  4.2  &  c  &  6  &  1800  &  19.350 $\pm$ 0.023 $\ast$  &  - \\
07/08/12  & 2456117.4 & 2.875 & 2.976 & 19.91 & 110.2 &  Cal      &  DLR-MKIII  &  0.314  &  2.8  &  4.2  &  c  &  7  &  2100  &  19.389 $\pm$ 0.038 $\ast$  &  - \\
07/10/12  & 2456118.8 & 2.884 & 3.002 & 19.76 & 110.4 &  UH2.2  &  Tek/WFGS2  &  0.340  &  1.7  &  6.0  &  p  &  2  &  1200  &  19.589 $\pm$ 0.014 $\ddagger$  &  -  \\
\tableline                              
\multicolumn{16} {c} {Supplemental Observations from Literature} \\                              
\tableline                              
UT Date  &  Julian Date  &  r  &  $\Delta$  &  $\alpha$  &  TA  &  Tel  &  Inst  &  Scale  &  G  &  RN  &  Sky  &  \#Obs  &  Exp  &  R Mag  &  Ref \\
\tableline                              
12/11/98  & 2451158.8 & 2.112 & 2.443 & 23.56 & 85.9 &  WHT  &  Tek 1024  &  0.110  &  1.8  &  4.7-6.5  &  p  &  1  &  300  &  18.60$\pm$0.03  & 1 \\
06/13/99  & 2451343.4 & 3.337 & 2.778 & 16.01 & 123.7 &  JKT  &  Tek 1024  &  0.330  &  ?  &  5  &  p  &  3  &  540  &  19.51$\pm$0.05  &  2\\
08/23/04$\ast$  & 2453241.0 & 2.348 & 1.538 & 18.31 & -95.7 &  CFHT  &  MegaCam  &  0.187  &  1.62  &  5  &  p  &  36  &  2160  &  -  &  3\\
11/29/04$\ast$  & 2453338.5 & 1.680 & 1.088 & 33.91 & -59.3 &  \textit{Spitzer}  & IRAC \footnote{Observations done in all four bands of 3.6 $\micron$, 4.5 $\micron$, 5.8 $\micron$ and 8 $\micron$ meaning a range of
  values for plate scale, gain, and read noise} &  $\sim$1.22  &  3.3-3.8  &  6.7-9.4  &  n/a  &  560  &  5824  &  -  &  4\\
12/05/04$\ast$  & 2453345.1 & 1.641 & 1.099 & 35.68 & -55.8 &  \textit{Spitzer}  &  MIPS  &  2.550  &  5  &  40  &  n/a  &  736  &  1832  &  -  &  5\\

\tableline
\vspace{0.5cm}
\end{tabular}
\end{table*}

%% file: tab-dust.tex
\begin{table}[h!t]
\tabletypesize{\small}
\centering
\caption{Dust Material Densities}
\label{tab:dust}
\begin{tabular}{l|c}
Material&Density (kg m$^{-3}$) \\
\tableline\tableline
Water Ice & 1000 \\
Olivine   & 3300 \\
Chondrite & 2200 \\
Silicate  & 2200 \\
Tholin    & 1450 \\
Magnetite & 5200 \\
Graphite  & 2200 \\
\end{tabular}
\end{table}

%% file: tab-FPResults.tex
\begin{table*}[h!t]
\tabletypesize{\small}
\centering
\caption{Summary of FP Modeling Results}
\label{tab:FPResult}
\begin{tabular}{c|c|c|c|c|c|c|c}
Observation &  Days from \footnote{Negative values are number of days before perihelion, Positive values are number of days after perihelion}                                      & n\footnote{From Equation \ref{equation:beta_n}} & Grain Sizes & Peak Activity & Duration of Activity & Grain Velocity & Notes \\
 Date &  Perihelion & & ($\micron$) & (Days)\footnote{Days since perihelion} & (Days) & (m s$^{-1}$) & \\
\tableline\tableline
1992\footnote{Two observations from January 5 and March 7} & 95, 156 & 3.5 - 5.0 & 40-400 & $\sim$0 & $>$157 & 1-13 & Tail feature only \\
Aug 23, 2004 &  -186 & 2.5 \& 3.8 & 1-8 & -206 \& -191 & $>$25 & 1-4 & Two jet outbursts \\
2012\footnote{Four observations from Feb 22, Mar 28, Apr 1, and Apr 15} &  126, 161, 165, 179 & 3.5 - 4.5 & 780-4000 & -25 to 0 & $\sim$200 & 0.5-2 & Tail Only \\
2012\footnote{Two observations from Mar 28 and Apr 1} & 161, 165 & 4.0 & 1.5-6.5 & 143 to 146 & 34 & 1-4 & Jet Only \\
\end{tabular}
\end{table*}

%% file: tab-sublimation.tex
\begin{table}[h!t]
\tabletypesize{\small}
\centering
\caption{Ice Sublimation Model Parameters} \label{tab:sublimation}
\begin{tabular}{lccc}
Parameters          & Value & Units          & Reference\\
\tableline\tableline
Nucleus             &       &                & \\
\tableline
Geometric Albedo              & 0.028 &                & \citet{millis1988} \\
Emissivity          & 0.9   &                & assumed \\
Phase Function      & 0.035 & mag deg$^{-1}$ & \citet{sekanina1974a} \\
Density             & 400   & kg m$^{-3}$    & \citet{Thomas2013} \\
Radius              & 4600  & m             & \citet{lowry2003} \\
\tableline\tableline
Dust                &       &                & \\
\tableline
Coma Phase Function & 0.02  & mag deg$^{-1}$ & 
\citet{Meech1987} \\
Density             & 2200  & kg m$^{-3}$    & Section 4.1.1 \\
Radius              & 3-8  & $\micron$       & Section 4.5.1\\
\end{tabular}
\end{table}

%% file: tab-apparition.tex
\begin{table}[h!t]
\tabletypesize{\small}
\centering
\caption{Apparition Dates} \label{tab:apparition}
\begin{tabular}{lllc}
ID &Range of Dates & Range of JD & \# of Obs \\
\tableline\tableline
1 & 07/04/1981 - 05/02/1988 & 2444789 - 2447283 & 5\footnote{four from Jewitt \& Meech (1985)} \\
2 & 05/03/1988 - 02/23/1995 & 2447284 - 2449771 & 10 \\ 
3 & 02/24/1995 - 11/02/2001 & 2449772 - 2452215 & 3 \\
4 & 11/03/2001 - 06/16/2008 & 2452216 - 2454633 & 0 \\
5 & 06/17/2008 - 03/03/2015 & 2454634 - 2457084 & 8 \\
6 & 03/04/2015 - 11/17/2021 & 2457085 - 2459535 & 0 \\
\end{tabular}
\end{table}

%% file: tab-fracactive.tex
\begin{table}[h!t]
\tabletypesize{\small}
\centering
\caption{Fractional Active Area of H$_2$O Model Results}
\label{tab:fracactive}
\begin{tabular}{l|c|c|c|c}
Size\footnote{Grain size} ($\micron$) & \multicolumn{4}{c}{Fractional Active Area of H$_2$O}
\\
\tableline\tableline
 & App 1 & App 2\footnote{We have a measured gas production rate
   during this apparition so the fractional active area is directly
   calculated so we conclusively use only 8 $\micron$ size grains} & App 3 & App 5 \\
\tableline
3 & 0.015 & - & 0.0016 & 0.0010  \\ 
4 & 0.020 & - & 0.0022 & 0.0015  \\
5 & 0.025 & - & 0.0027 & 0.0018  \\
6 & 0.030 & - & 0.0035 & 0.0022  \\
7 & 0.040 & - & 0.0040 & 0.0025  \\ 
8 & 0.050 & 0.0063 & 0.0045 & 0.0028  \\ 
\end{tabular}
\end{table}